\documentclass[cameraready]{Interspeech}

\title{More than a feeling: Expressive style influences cortical speech tracking in subjective cognitive decline}

\author[affiliation={1}, orcid=0000-0001-9194-2498, correspondingauthor]{Matthew King-Hang}{Ma}
\author[affiliation={1}, equalcontribution]{Yun}{Feng}
\author[affiliation={1}, equalcontribution]{Cloris Pui-Hang}{Li}
\author[affiliation={1,2}, orcid=0000-0002-2486-2268]{Manson Cheuk-Man}{Fong}

\address{
    $^1$ Research Centre for Language, Cognition, and Neuroscience, Department of Language Science and Technology, The Hong Kong Polytechnic University, Hong Kong \\
    $^2$ Research Institute for Smart Ageing, The Hong Kong Polytechnic University, Hong Kong
}

\email{khmma@polyu.edu.hk, yunfeng@poly.edu.hk, ph2li@polyu.edu.hk, cmmfong@polyu.edu.hk}


\usepackage{comment}
\usepackage{booktabs, tabularx, multirow}

\begin{document}

\maketitle

\begin{abstract}
Subjective cognitive decline (SCD) doubles dementia risk. This study investigates how self-perceived cognitive worsening shapes neural dynamics during naturalistic speech perception. EEG was collected from 60 cognitively normal older adults as they listened to speech varied in prosodic contexts, categorized by expressive style (scrambled, descriptive, dialogue, exciting). Encoding models mapping three speech representations—acoustic, subsyllabic segmentation and phonotactic features—to ongoing EEG signals were built. Cortical tracking strength (CTS) showed that models fitted with subsyllabic linguistic features outperformed acoustic ones. Crucially, greater SCD severity was associated with weaker CTS of (1) subsyllabic linguistic but not acoustic features, and (2) prosodically flat speech (scrambled and descriptive). Thus, the CTS of higher-level linguistic features while listening to prosodically flat speech may serve as a potential neural marker for early-stage cognitive decline.
\end{abstract}

\section{Introduction}
\begin{figure*}[!t]
    \centering
    \includegraphics[width=\linewidth]{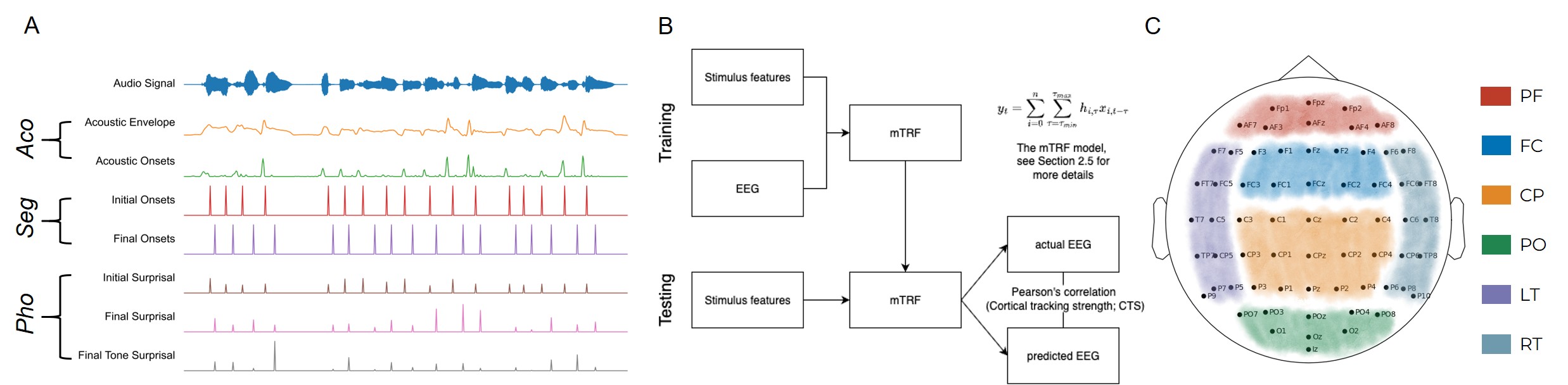}
    \caption{Methodology. \textbf{A.} Three different set of stimulus features (Aco: Acoustic, Seg: Segmentation, Pho: Phonotactic); \textbf{B.} Training and testing of mTRF models; and the derivation of cortical tracking strength. \textbf{C.} The grouping of 64 electrodes into 6 sites.}
    \label{fig:mtrf_framework}
\end{figure*}

\label{sec:intro}
Population ageing worldwide leads to the increasing prevalence of cognitive decline, resulting in a pressing need for effective monitoring and early intervention strategies \cite{Hong2023ImpactsBaselineBiomarkers}. Subjective cognitive decline (SCD), a self-perceived cognitive worsening despite normal objective cognitive test performance, has emerged as an early indicator of potential impairment \cite{Jessen2020CharacterisationSubjectiveCognitive}. Evidence suggests SCD doubles the risk of progressing to mild cognitive impairment (MCI) and dementia, particularly for those reporting complaints at a younger age (around 60) \cite{Jessen2020CharacterisationSubjectiveCognitive}. While most studies focused on memory decline \cite{Jessen2020CharacterisationSubjectiveCognitive}, everyday functioning relies on broader cognitive abilities. Exploring neural correlates of SCD during ecologically valid tasks, such as naturalistic speech perception, is therefore central to understanding cognitive decline comprehensively and informing early detection strategies.

Effective speech comprehension is essential for older adults' social connection and quality of life \cite{Heinrich2016EffectiveCommunicationFundamental}, yet it often declines with age. This stems from degrading auditory perception and cognitive abilities, such as inhibitory control \cite{Fong2021CanInhibitionDeficit}. The degradation of auditory perception, beginning as early as middle age \cite{Helfer2017AgeRelatedChangesObjective}, impairs both incoming phonological representations and the perception of speech prosody—--the vocal modulations of suprasegmental information---that is fundamental to social interaction \cite{Mixdorff2017AudiovisualExpressionsAttitude}. To compensate for degraded bottom-up acoustic representations, speech comprehension relies increasingly on cognitively demanding top-down processing. For example, older adults rely more heavily on semantic rather than acoustic cues during prosody perception \cite{Ben-David2019AgeRelatedDifferencesPerception}.

Pioneering studies have discovered that ongoing neural signals, such as electroencephalography (EEG), can track incoming auditory stimuli \cite{Ahissar2001SpeechComprehensionCorrelateda, Ding2012NeuralCodingContinuous}. Using encoding models like the multivariate temporal response function (mTRF), researchers have shown that cortical tracking is not merely an epiphenomenon. Instead, it facilitates general auditory processing \cite{Lalor2025GenerativeMechanismsUnderlying, Bolt2024NeuralEncodingLinguistic} alongside speech-specific processing at the phoneme \cite{Prinsloo2022GeneralAuditorySpeechSpecific}, word \cite{Mesik2021EffectsAgeCortical}, and prosodic levels \cite{Bernardo2024NeuralTrackingProsodic, Antonicelli2024CorticalTrackingProsody}. Furthermore, cortical tracking is modulated by prior context \cite{DiLiberto2018CorticalMeasuresPhonemeLevel} and intelligibility; for instance, phoneme feature tracking is only significant when speech is understood \cite{Prinsloo2022GeneralAuditorySpeechSpecific}. These findings demonstrate cortical tracking's functional and predictive role in speech comprehension.

Cortical tracking studies in older adults show enhanced tracking of higher-level linguistic features \cite{Mesik2021EffectsAgeCortical} and reduced tracking of acoustic features \cite{Herrmann2022NeuralSignatureRegularity}, a pattern more pronounced with poorer inhibition abilities \cite{Decruy2019EvidenceEnhancedNeural}. These results suggest that the tracking of linguistic features compensates for degraded acoustic and cognitive functions. Furthermore, tracking of the speech envelope provides unique contributions by aiding in temporal prediction \cite{Bernardo2024NeuralTrackingProsodic, Lamekina2024SpeechProsodyServes}, though this ability is shown to decline with age \cite{Gillis2023NeuralTrackingLinguistic}. While cortical tracking findings in cognitively declined populations remain scarce and inconsistent \cite{Bolt2024NeuralEncodingLinguistic, Bidelman2017MildCognitiveImpairment}, behavioral studies clearly show these populations exhibit significant impairments in recognizing emotional prosody \cite{Amlerova2022EmotionalProsodyRecognition, Jiang2024ComprehensionAcousticallyDegraded}. Examining how the cortical tracking of diverse speech features changes in individuals with SCD across varying prosodic contexts, could therefore reveal novel, early-stage candidate neural markers for cognitive decline and provide insight into the underlying neural changes.

The present study utilizes cortical tracking to investigate how self-perceived worsening of cognition affects speech perception. We specifically tapped into features across two levels of speech processing: acoustic and subsyllabic. While tracking of acoustic features reflects bottom-up sensory processing, tracking of subsyllabic segmentation and phonotactic features captures higher-level linguistic processing. Including both features allows us to model the tracking of subsyllabic structures, both independently and alongside their statistical predictions. Furthermore, we evaluate the impact of varying prosodic contexts using custom stimuli categorized by expressive style.

We hypothesize that (1) cortical tracking strength (CTS) for higher-level subsyllabic features will be significantly modulated by SCD severity, whereas (2) CTS for bottom-up acoustic features will remain relatively stable, reflecting a preservation of sensory processing relative to higher-level linguistic processing. 
Regarding the impact of prosodic contexts, we hypothesize that (3) CTS for prosodically rich stimuli will be more sensitive to SCD severity. While acoustic processing under clean speech condition may be preserved,
the need to integrate rapidly changing prosodic cues and multi
talker dynamics imposes higher listening demands \cite{Kovacs2023SpeechProsodySupports}.

\section{Materials and Methods}
\subsection{Participants}
A total of 62 (30F) Cantonese speakers with no known neurological disorders was recruited. These individuals were aged 60 -- 70 ($M$ = 65.2, $SD$ = 3.0). Each participant was cognitively normal per MoCA-HK, adjusted for education years \cite{Wong2015MontrealCognitiveAssessment}. Degree of SCD was assessed using the 14-item Subjective Cognitive Decline Scale (SCDS) \cite{Tsai2021DevelopmentSubjectiveCognitive}. SCDS scores ranged from 14 to 58 out of 60 ($M$ = 34.6, $SD$ = 11.3). Hearing thresholds were measured by Pure-Tone Audiometry (PTA). Written informed consent was obtained from all participants. 

\subsection{Stimuli and experimental procedures}
Speech stimuli of four expressive styles were constructed: scrambled, descriptive, dialogue, and exciting. The scrambled and descriptive style stimuli were synthesized with \textit{Amazon Polly} from transcribed radio weather reports; the descriptive style used the original text, while the scrambled style used a word-level scrambled version. Dialogue-style stimuli were sourced from a local community radio program, and exciting-style stimuli were curated from a radio drama. While the scrambled and descriptive styles consist of single-talker speech, the dialogue and exciting styles include two or more talkers. Including multi-talker scenario was a methodological choice to enhance ecological validity, as well as to create cognitively challenging conditions that demand both sensory and cognitive resources. These complex scenarios open up the potential to reveal subtle SCD-induced neural changes, which could otherwise be masked while listening to simple, single-talker speech.

From an initial set of eight stimuli per style, we recruited an independent group of 24 (12F) raters (age $= 22.0 \pm 2.8$) to evaluate the perceived valence, arousal, and dominance of these recordings. Then, we conducted K-means clustering ($k=4$) on the 3D Valence-Arousal-Dominance (VAD) space. For each style, we selected four out of eight stimuli that were closest to their respective cluster center for the subsequent EEG study (average ratings in Table \ref{tab:averaged_ratings}). The standard deviation (SD) of the fundamental frequency (F0) and root-mean-square (RMS) amplitude of selected stimuli were also computed. As shown in Table 1, the dialogue and exciting styles exhibited higher F0 or intensity SD, revealing their richer prosodic nature.

\begin{table}[!hbt]
\centering
\caption{Mean stimulus ratings and acoustic properties. Val: Valence; Aro: Arousal; Dom: Dominance. $\text{F0}_\text{SD}$: SD of F0 in Hz. $\text{RMS}_\text{SD}$: SD of RMS amplitude.}
\label{tab:averaged_ratings}
\begin{tabularx}{\linewidth}{lXXXXX}
  \toprule
 Style & Val & Aro & Dom & $\text{F0}_\text{SD}$ & $\text{RMS}_\text{SD}$\\
  \midrule
    scrambled & 2.00 & 1.69 & 2.41 & 48.53 & 5.40\\ 
    descriptive & 2.57 & 1.94 & 2.69 & 46.75 & 5.42\\ 
    dialogue & 3.61 & 3.31 & 3.10 & 67.34 & 5.31\\ 
    exciting & 3.60 & 3.84 & 3.38 & 81.67 & 6.12\\
   \bottomrule
\end{tabularx}
\end{table}
\subsection{Data acquisition and preprocessing}
EEG data were recorded using a 64-channel BioSemi ActiveTwo system at a 2048 Hz sampling rate. Eye movements were monitored with horizontal and vertical electrooculograms. Participants sat approximately 70 cm from a monitor while auditory stimuli were presented via two speakers. During the recording, they listened to 16 one-minute recordings of clean speech (4 per expressive style) in a random order. After each recording, they used a five-point Likert scale to rate perceived valence, arousal, and dominance with the Self-Assessment Manikin \cite{Bradley1994MeasuringEmotionSelfAssessment}, as well as their overall enjoyment. Two participants with unusually low variability in their behavioral ratings ($Z \leq -1.96$ for standard deviation of valence, arousal, or dominance) were excluded, leaving 60 (30F) participants.

Prior to any preprocessing steps, the raw EEG signals were visually inspected to exclude channels contaminated by body movements. The signals were first downsampled to 512 Hz and average re-referenced. Independent component analysis (ICA) was applied to remove eye artifacts following a z-score thresholding approach \cite{Ma2021RegularityRandomnessAgeing}. Afterwards, channels that were excluded during visual inspection were interpolated. The signals were then band-pass filtered between 1 and 40 Hz using a third-order Butterworth filter. The above steps were conducted with custom Python scripts with \textit{MNE-Python} \cite{Gramfort2013MEGEEGData}.

\subsection{Feature extraction}
\subsubsection{Acoustic features}
\label{sec:acoustic-features}
The speech envelope and its onset were included as acoustic features following existing studies \cite{ Bolt2024NeuralEncodingLinguistic, Prinsloo2022GeneralAuditorySpeechSpecific, Bernardo2024NeuralTrackingProsodic, DiLiberto2018CorticalMeasuresPhonemeLevel, Issa2024SpeechEnvelopeCortical}. To extract the speech envelope, each audio signal was first downsampled to 128 Hz, and then filtered with 24 gammatone filters ranging from 50 Hz to 8000 Hz. The filtered responses were then scaled to the power of 0.6 to approximate the nonlinear relationship between sound intensity and perceived loudness \cite{Issa2024SpeechEnvelopeCortical}. These 24 filtered signals were then averaged and filtered from 0.5 Hz to 25 Hz to obtain the final envelope. The envelope onset was computed as the first-order differences of the speech envelope.

\subsubsection{Subsyllabic segmentation and phonotactic features}
\label{sec:subsyllabic-features}
Subsyllabic segmentation was obtained by training a Cantonese Montreal Forced Aligner (MFA) on the 109-hour Common Voice Hong Kong Chinese Corpus (zh-HK 21.0), which features 3,084 speakers \cite{So2025PerformanceMontrealForced}. A pronunciation dictionary was generated with \textit{CharsiuG2P} \cite{Zhu2022ByT5ModelMassively} grapheme-to-phoneme conversion tool. The resulting MFA alignments were then manually refined in Praat \cite{Boersma2024PraatDoingPhonetics}. Following standard Chinese phonological analysis, the onset times of initials and finals rather than phonemes were extracted as segmentation features \cite{Wang1973ChineseLanguage}. 

To the best of our knowledge, there are no publicly available Cantonese subsyllabic probabilistic phonotactic features. We therefore extracted them by deriving the initial, final and tone surprisal values from the Hong Kong Cantonese Corpus, which contains around 230,000 words \cite{Wong2019HongKongCantonese}. The texts were parsed with \textit{pycantonese} \cite{Lee2022PyCantoneseCantoneseLinguistics}, and surprisal values were calculated as the negative logarithm of the probability for each component: unigram probability for initials $P(\text{initial})$, conditional probability for finals $P(\text{final}|\text{initial})$, and conditional probability for tones $P(\text{tone}|\text{initial,final})$. All subsyllabic segmentation and phonotactic features were downsampled to 128 Hz. These features are now available at \cite{neurothew_2026_20748010} to facilitate future research.

\subsection{Multivariate temporal response function (mTRF)}
To investigate the cortical tracking of speech, we modeled the relationship between the EEG signals (dependent variable, $y$) and various stimulus features (predictor variables, $x$) using a multivariate temporal response function (mTRF). It models the EEG signals as a linear convolution of the stimulus features with a kernel $h$ (see also Figure \ref{fig:mtrf_framework}B):
\begin{equation}
    \label{eqn:convolution}
    y_t = \sum_{i=0}^n \sum_{\tau=\tau_{min}}^{\tau_{max}} h_{i, \tau}x_{i, t-\tau}    
\end{equation}
where $y_t$ is the EEG signal at one channel at time $t$, $n$ is the total number of predictor variables, and $x_i$ is the $i$-th predictor. We fitted all mTRF models using the Boosting algorithm implemented in the \textit{Eelbrain} Python toolkit \cite{Brodbeck2023EelbrainPythonToolkit}, which can yield more accurate mTRF estimates than conventional ridge regression. Prior to fitting, the preprocessed EEG signals were downsampled to 128 Hz and low-pass filtered at 25 Hz to match the sampling rate of the stimulus features. The mTRFs were estimated over time lags ranging from -200 ms to 600 ms. To investigate the tracking of different stimulus features, we fitted three models using distinct feature sets: acoustic (\textit{Aco}), segmentation (\textit{Seg}), and phonotactic (\textit{Pho}) features (Figure \ref{fig:mtrf_framework}A). This procedure was repeated for each of the four expressive styles, yielding a total of 12 models.

Model performance was evaluated using a leave-one-out cross-validation scheme. Within $k$ stimuli, a model was trained on data from $k - 1$ stimuli and then tested on the held-out stimulus. This process was repeated $k$ times, with each stimulus serving as the test set once. CTS was quantified as the Pearson's correlation coefficient between the observed and predicted EEG signals, averaged across all $k$ folds. 

\subsection{Statistical analyses}
For statistical analysis, the 64 channels were grouped into 6 scalp sites (Figure \ref{fig:mtrf_framework}C): prefrontal (FP); frontocentral (FC); centroparietal (CP); parieto-occipital (PO); left temporal (LT) and right temporal (RT) \cite{Ma2021RegularityRandomnessAgeing}. The maximum CTS per site served as the dependent variable to capture the strongest local tracking response while reducing dilution from weakly responsive electrodes. To examine how CTS is modulated by SCD severity, mTRF models and expressive styles, we fitted a linear mixed effect model in \textit{R} with \textit{lme4} \cite{Bates2014FittingLinearMixedEffects} as in Equation \ref{eq:lme_cts}:
\begin{equation}
\label{eq:lme_cts}
\begin{split}
\text{CTS} \sim &  \ \text{SCDS}*\text{Site}*\text{Model}*\text{ExpressStyle} + \text{Age} + \text{Gender} \ + \\
&\text{EducationYears} + \text{MoCA} + \text{PTA} + \text{Valence}\ + \\
&\text{Arousal} + \text{Dominance} + \text{Enjoy} + (1|\text{Participant})
\end{split}
\end{equation}
where $\text{SCDS}$ refers to the SCDS scores, $\text{Model}$ refers to mTRF models fitted with three distinct feature sets (\textit{Aco}, \textit{Seg}, \textit{Pho}), $\text{Site}$ refers to the six scalp sites. Covariates included age, gender, education years, MoCA-HK score, PTA, perceived valence, arousal, dominance and enjoyment, along with the participant-specific random intercepts. PTA was defined as the average of the hearing thresholds at 500, 1000, 2000 and 4000 Hz \cite{Bolt2024NeuralEncodingLinguistic}.

The full model in Equation \ref{eq:lme_cts} was then reduced to a simpler model with backward elimination. Type-III ANOVA was then performed, and all post-hoc comparisons were conducted with estimated marginal means using \textit{emmeans} \cite{Lenth2025EmmeansEstimatedMarginal}.

\begin{figure*}[!t]
    \centering
    \includegraphics[width=\linewidth]{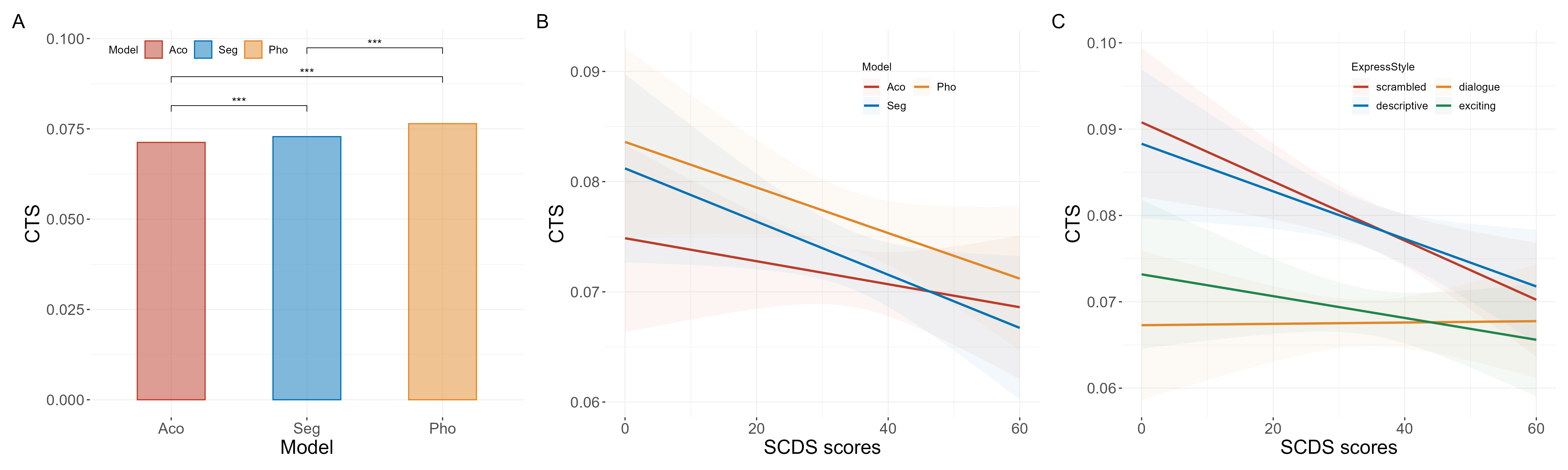}
    \caption{\textbf{A.} CTS across models; \textbf{B.} SCDS $\times$ Model interaction. \textbf{C.} SCDS $\times$ ExpressStyle interaction. Y-axis: estimated marginal means; ribbons: 95\% CIs.} 
    \label{fig:plot_combined_results}
\end{figure*}
\section{Results}

The final reduced model is shown in Equation \ref{eq:lme_cts_final}.
\begin{equation}
\label{eq:lme_cts_final}
\begin{split}
\text{CTS} \sim &  \ \text{SCDS} + \text{Site} + \text{Model} + \text{ExpressStyle} + \text{Arousal} \ + \\
& \text{SCDS:Site}\ + \text{SCDS:Model} + \text{Site:Model}\ + \\
& \text{SCDS:ExpressStyle} + \text{Site:ExpressStyle}\ + \\ & \text{ExpressStyle:Model}\ + (1|\text{Participant})
\end{split}
\end{equation}
As Equation \ref{eq:lme_cts_final} shows, none of the three-way or four-way interactions survived after backward elimination. Among all significant factors, we were mostly interested in the Model main effect, the interaction between SCDS scores and expressive style ($F(3, 4208.67)=25.43, p <.001$) and that between SCDS scores and Model ($F(2, 4208.02)=7.07, p <.001$), addressing our three predictions. Other effects fall beyond our scope.

\subsection{Subsyllabic models outperform the acoustic model}
A significant main effect of mTRF models was found ($F(2, 4208.02)=21.48, p <.001$). Post-hoc comparisons (Figure \ref{fig:plot_combined_results}A) revealed a hierarchy in model performance. The \textit{Pho} model significantly outperformed the \textit{Seg} model, which in turn demonstrated significantly higher CTS compared to the \textit{Aco} model ($ps<.001$ for all pairs).

\subsection{SCD modulates CTS in linguistic (not acoustic) models}
To decompose the significant two-way interaction between SCDS scores and mTRF models, we first conducted a post-hoc simple slope analysis and found SCDS scores significantly modulated CTS negatively for the \textit{Seg} model ($t(62.26)=-2.05, p = .044$) while the trend for \textit{Pho} model was marginally significant ($t(62.26)=-1.76, p = .084$).

Subsequent pairwise comparisons were conducted with Figure \ref{fig:plot_combined_results}B showing the interaction plot. The trends of both the \textit{Seg} ($t(4208)=-3.61, p = .001$) and \textit{Pho} ($t(4208)=-2.70, p =.019$) models were significantly more negative than that of the \textit{Aco} model, while there were no significant differences between \textit{Seg} and \textit{Pho} models.

\subsection{SCD modulates CTS in flat (not rich) styles}
To decompose the significant two-way interaction between SCDS scores and expressive style, we first conducted a post-hoc simple slope analysis and found SCDS significantly modulated CTS negatively for scrambled- ($t(64.37)=-2.89, p =.005$) and descriptive-style speech ($t(64.44)=-2.32, p =.023$).

Subsequent pairwise comparisons were conducted with Figure \ref{fig:plot_combined_results}C showing the interaction plot. The trend for scrambled-style was significantly more negative than that of the dialogue- ($t(4209.41)=-7.97, p <.001$) and exciting-style ($t(4209.53)=-4.92, p <.001$). Similarly, the trend for descriptive-style was also significantly more negative than that of the dialogue-style ($t(4208.09)=-6.47, p <.001$) and the exciting-style ($t(4208.12)=-3.41, p = .004$).

\section{Discussion}
The present study investigated how CTS is modulated by SCD severity, different mTRF models (trained by different acoustic or subsyllabic linguistic features), and expressive styles. We first showed that the phonotactic model outperformed both the segmentation and acoustic models in terms of CTS. Thus, higher-level linguistic information better explained EEG signals during speech perception in cognitively normal older adults. Consistent with previous findings \cite{Mesik2021EffectsAgeCortical, Bolt2024HearingCognitiveDecline}, this confirms that successful speech perception in aging relies more heavily on top-down linguistic processing. Establishing this baseline sets the stage to examine how SCD disrupts these processes. 

Supporting our first two hypotheses, we found that SCD modulates cortical tracking differently depending on the level of speech processing. As predicted, greater SCD severity was associated with weaker cortical tracking of higher-level, subsyllabic linguistic (segmentation and phonotactics) features, but not acoustic features. It suggests that SCD may specifically affect linguistic features, reflecting early vulnerability in processing linguistic information that support speech perception.

We initially predicted that CTS for prosodically rich (dialogue and exciting), but not flat (scrambled and descriptive), stimuli would be most negatively impacted by SCD severity. This is due to the need to integrate rapidly changing pitch, intensity, duration, and rhythm cues, as well as multi-talker dynamics. High demands are placed on both the sensory and cognitive systems in older listeners, whether they are cognitively healthy or impaired \cite{Bidelman2017MildCognitiveImpairment}. Yet, our results showed the opposite pattern. A possible explanation is that despite the local acoustic complexity, rich prosody may be more beneficial than detrimental at the SCD stage. Existing studies have shown that prosodic cues may act as reliable acoustic and perceptual scaffolds in aiding temporal predictions \cite{Lamekina2024SpeechProsodyServes} and syntactic parsing \cite{Degano2024SpeechProsodyEnhances}. Thus, our results showed that, given the specific materials, the availability of prosodic cues may allow even participants with greater subjective concerns to compensate for weaker speech tracking. In contrast, flat speech provides fewer prosodic cues, requiring greater reliance on subsyllabic segmentation and phonotactic probabilities. Together with the observed SCD-related reduction in the tracking of linguistic features, our findings therefore suggest that higher-level linguistic processing, when unsupported by prosodic scaffolding, is most vulnerable to SCD.

Our study is limited in that only two levels of speech processing were modeled. Future studies should examine features including but not limited to word-level (e.g., word surprisal \cite{Goldstein2022SharedComputationalPrinciples}) and explicit prosodic-level (e.g., pitch contour \cite{Gnanateja2025CorticalProcessingDiscrete}) features to gain a more comprehensive understanding of declining linguistic processing abilities and to test our speculated scaffolding role of prosody. Additionally, leveraging the EEG oscillatory hierarchy \cite{Lalor2025GenerativeMechanismsUnderlying} could help localize specific linguistic deficits and corresponding brain rhythms.

\section{Conclusion}
Utilizing speech stimuli across varying prosodic contexts, we investigated how subjective cognitive decline (SCD) impacts the cortical tracking of acoustic, subsyllabic segmentation and phonotactic features. We found that greater subjective cognitive concern is associated with weaker cortical tracking of (1) higher-level, subsyllabic linguistic features, and (2) prosodically flat speech. These results suggest that individuals with greater subjective cognitive concerns may make use of rich prosodic cues to compensate for their diminished higher-level linguistic processing. Consequently, the cortical tracking of prosodically flat speech could serve as a candidate neural marker for early-stage cognitive decline.

\section{Acknowledgements}
This work was supported by the HKRGC Postdoctoral Fellowship Scheme awarded to M.K-H.M.. We thank the
University Research Facility for Big Data Analytics
(UBDA), HKPolyU, for making the GPU virtual
machines available for the computations involved.

\section{Generative AI Use Disclosure}
Gemini 3.1 Pro was used to improve the grammar and syntax after the author had written the draft. All AI-generated suggestions were manually reviewed. 

\bibliographystyle{IEEEtran}
\bibliography{main.bib}

@inproceedings{Antonicelli2024CorticalTrackingProsody,
  title = {Cortical Tracking of Prosody after Stroke and in Aging: Evidence from Magnetoencephalography},
  shorttitle = {Cortical Tracking of Prosody after Stroke and in Aging},
  booktitle = {Speech {{Prosody}} 2024},
  author = {Antonicelli, Giada and Molinaro, Nicola and Riva, Patricia De La and Laspiur, Raquel and Turiso, Arantza Lopez De and Mancini, Simona},
  year = {2024},
  month = jul,
  pages = {21--26},
  publisher = {ISCA},
  doi = {10.21437/SpeechProsody.2024-5},
  urldate = {2025-02-13},
  langid = {english}
}

@inproceedings{Bernardo2024NeuralTrackingProsodic,
  title = {Neural Tracking of Prosodic Structure in Delexicalized Speech},
  booktitle = {Speech {{Prosody}} 2024},
  author = {Bernardo, Andr{\'e} and Correia, Pedro and Vig{\'a}rio, Marina and Vig{\'a}rio, Ricardo and Frota, S{\'o}nia},
  year = {2024},
  month = jul,
  pages = {1140--1144},
  publisher = {ISCA},
  doi = {10.21437/SpeechProsody.2024-230},
  urldate = {2025-02-13},
  langid = {english}
}

@article{Bolt2024NeuralEncodingLinguistic,
  title = {Neural Encoding of Linguistic Speech Cues Is Unaffected by Cognitive Decline, but Decreases with Increasing Hearing Impairment},
  author = {Bolt, Elena and Giroud, Nathalie},
  year = {2024},
  month = aug,
  journal = {Scientific Reports},
  volume = {14},
  number = {1},
  pages = {19105},
  publisher = {Nature Publishing Group},
  issn = {2045-2322},
  doi = {10.1038/s41598-024-69602-1},
  urldate = {2024-09-01},
  copyright = {2024 The Author(s)},
  langid = {english}
}

@article{Brodbeck2023EelbrainPythonToolkit,
  title = {Eelbrain, a {{Python}} Toolkit for Time-Continuous Analysis with Temporal Response Functions},
  author = {Brodbeck, Christian and Das, Proloy and Gillis, Marlies and Kulasingham, Joshua P and Bhattasali, Shohini and Gaston, Phoebe and Resnik, Philip and Simon, Jonathan Z},
  year = {2023},
  month = nov,
  journal = {Elife},
  volume = {12},
  issn = {2050-084X},
  doi = {10.7554/eLife.85012}
}

@article{Issa2024SpeechEnvelopeCortical,
  title = {On the Speech Envelope in the Cortical Tracking of Speech},
  author = {Issa, Mohamed F. and Khan, Izhar and Ruzzoli, Manuela and Molinaro, Nicola and Lizarazu, Mikel},
  year = {2024},
  month = aug,
  journal = {NeuroImage},
  volume = {297},
  pages = {120675},
  issn = {1053-8119},
  doi = {10.1016/j.neuroimage.2024.120675},
  urldate = {2024-12-07}
}

@article{Jessen2020CharacterisationSubjectiveCognitive,
  title = {The Characterisation of Subjective Cognitive Decline},
  author = {Jessen, Frank and Amariglio, Rebecca E and Buckley, Rachel F and {van der Flier}, Wiesje M and Han, Ying and Molinuevo, Jos{\'e} Luis and Rabin, Laura and Rentz, Dorene M and {Rodriguez-Gomez}, Octavio and Saykin, Andrew J and Sikkes, Sietske A M and Smart, Colette M and Wolfsgruber, Steffen and Wagner, Michael},
  year = {2020},
  month = mar,
  journal = {Lancet Neurology},
  volume = {19},
  number = {3},
  pages = {271--278},
  issn = {1474-4422},
  doi = {10.1016/S1474-4422(19)30368-0}
}

@article{Tsai2021DevelopmentSubjectiveCognitive,
  title = {Development of the {{Subjective Cognitive Decline Scale}} for {{Mandarin-Speaking Population}}},
  author = {Tsai, Hsing-Fang and Wu, Chi-Hsun and Hsu, Chih-Cheng and Liu, Chien-Liang and Hsu, Yen-Hsuan},
  year = {2021},
  journal = {Am. J. Alzheimers. Dis. Other Demen.},
  volume = {36},
  pages = {15333175211038237},
  issn = {1533-3175},
  doi = {10.1177/15333175211038237}
}

@article{Wong2015MontrealCognitiveAssessment,
  title = {Montreal {{Cognitive Assessment}}: {{One Cutoff Never Fits All}}},
  shorttitle = {Montreal {{Cognitive Assessment}}},
  author = {Wong, Adrian and Law, Lorraine S. N. and Liu, Wenyan and Wang, Zhaolu and Lo, Eugene S. K. and Lau, Alexander and Wong, Lawrence K. S. and Mok, Vincent C. T.},
  year = {2015},
  month = dec,
  journal = {Stroke},
  volume = {46},
  number = {12},
  pages = {3547--3550},
  issn = {1524-4628},
  doi = {10.1161/STROKEAHA.115.011226},
  langid = {english},
  pmid = {26470772}
}

@article{Gramfort2013MEGEEGData,
  title = {{{MEG}} and {{EEG}} Data Analysis with {{MNE-Python}}},
  author = {Gramfort, Alexandre and Luessi, Martin and Larson, Eric and Engemann, Denis A and Strohmeier, Daniel and Brodbeck, Christian and Goj, Roman and Jas, Mainak and Brooks, Teon and Parkkonen, Lauri and H{\"a}m{\"a}l{\"a}inen, Matti},
  year = {2013},
  month = dec,
  journal = {Frontiers in Neuroscience},
  volume = {7},
  pages = {267},
  issn = {1662-4548},
  doi = {10.3389/fnins.2013.00267}
}

@article{Bradley1994MeasuringEmotionSelfAssessment,
  title = {Measuring Emotion: The {{Self-Assessment Manikin}} and the {{Semantic Differential}}},
  author = {Bradley, M M and Lang, P J},
  year = {1994},
  month = mar,
  journal = {J. Behav. Ther. Exp. Psychiatry},
  volume = {25},
  number = {1},
  pages = {49--59},
  publisher = {Elsevier},
  issn = {0005-7916},
  doi = {10.1016/0005-7916(94)90063-9}
}

@article{Hong2023ImpactsBaselineBiomarkers,
  title = {Impacts of Baseline Biomarkers on Cognitive Trajectories in Subjective Cognitive Decline: The {{CoSCo}} Prospective Cohort Study},
  shorttitle = {Impacts of Baseline Biomarkers on Cognitive Trajectories in Subjective Cognitive Decline},
  author = {Hong, Yun Jeong and Ho, SeongHee and Jeong, Jee Hyang and Park, Kee Hyung and Kim, SangYun and Wang, Min Jeong and Choi, Seong Hye and Yang, Dong Won and {CoSCo study group}},
  year = {2023},
  month = aug,
  journal = {Alzheimer's Research \& Therapy},
  volume = {15},
  number = {1},
  pages = {132},
  issn = {1758-9193},
  doi = {10.1186/s13195-023-01273-y},
  urldate = {2025-02-20},
  langid = {english}
}

@misc{Boersma2024PraatDoingPhonetics,
  title = {Praat: Doing Phonetics by Computer},
  author = {Boersma, Paul and Weenink, David},
  year = {2024},
  month = jun
}

@inproceedings{Zhu2022ByT5ModelMassively,
  title = {{{ByT5}} Model for Massively Multilingual Grapheme-to-Phoneme Conversion},
  booktitle = {Proc. {{Interspeech}} 2022},
  author = {Zhu, Jian and Zhang, Cong and Jurgens, David},
  year = {2022},
  pages = {446--450},
  doi = {10.21437/Interspeech.2022-538},
  urldate = {2025-09-09}
}

@misc{Lalor2025GenerativeMechanismsUnderlying,
  title = {On the Generative Mechanisms Underlying the Cortical Tracking of Natural Speech: A Position Paper},
  author = {Lalor, Edmund C. and Nidiffer, Aaron R.},
  year = {2025}
}

@article{Prinsloo2022GeneralAuditorySpeechSpecific,
  title = {General {{Auditory}} and {{Speech-Specific Contributions}} to {{Cortical Envelope Tracking Revealed Using Auditory Chimeras}}},
  author = {Prinsloo, Kevin D. and Lalor, Edmund C.},
  year = {2022},
  month = oct,
  journal = {Journal of Neuroscience},
  volume = {42},
  number = {41},
  pages = {7782--7798},
  publisher = {Society for Neuroscience},
  issn = {0270-6474, 1529-2401},
  doi = {10.1523/JNEUROSCI.2735-20.2022},
  urldate = {2025-09-10},
  chapter = {Research Articles},
  copyright = {Copyright {\copyright} 2022 Prinsloo and Lalor. This is an open-access article distributed under the terms of the Creative Commons Attribution 4.0 International license, which permits unrestricted use, distribution and reproduction in any medium provided that the original work is properly attributed.},
  langid = {english},
  pmid = {36041853}
}

@article{Decruy2019EvidenceEnhancedNeural,
  title = {Evidence for Enhanced Neural Tracking of the Speech Envelope Underlying Age-Related Speech-in-Noise Difficulties},
  author = {Decruy, Lien and Vanthornhout, Jonas and Francart, Tom},
  year = {2019},
  month = aug,
  journal = {J. Neurophysiol.},
  volume = {122},
  number = {2},
  pages = {601--615},
  issn = {0022-3077},
  doi = {10.1152/jn.00687.2018}
}

@article{Lamekina2024SpeechProsodyServes,
  title = {Speech {{Prosody Serves Temporal Prediction}} of {{Language}} via {{Contextual Entrainment}}},
  author = {Lamekina, Yulia and Titone, Lorenzo and Maess, Burkhard and Meyer, Lars},
  year = {2024},
  month = jul,
  journal = {Journal of Neuroscience},
  volume = {44},
  number = {28},
  publisher = {Society for Neuroscience},
  issn = {0270-6474, 1529-2401},
  doi = {10.1523/JNEUROSCI.1041-23.2024},
  urldate = {2025-01-02},
  chapter = {Research Articles},
  copyright = {Copyright {\copyright} 2024 Lamekina et al.. This is an open-access article distributed under the terms of the Creative Commons Attribution 4.0 International license, which permits unrestricted use, distribution and reproduction in any medium provided that the original work is properly attributed.},
  langid = {english},
  pmid = {38839302}
}

@article{Ben-David2019AgeRelatedDifferencesPerception,
  title = {Age-{{Related Differences}} in the {{Perception}} of {{Emotion}} in {{Spoken Language}}: {{The Relative Roles}} of {{Prosody}} and {{Semantics}}},
  shorttitle = {Age-{{Related Differences}} in the {{Perception}} of {{Emotion}} in {{Spoken Language}}},
  author = {{Ben-David}, Boaz M. and {Gal-Rosenblum}, Sarah and {van Lieshout}, Pascal H. H. M. and Shakuf, Vered},
  year = {2019},
  month = apr,
  journal = {Journal of Speech, Language and Hearing Research (Online)},
  volume = {62},
  number = {4S},
  pages = {1188--1202},
  publisher = {American Speech-Language-Hearing Association},
  address = {Rockville, United States},
  doi = {10.1044/2018_JSLHR-H-ASCC7-18-0166},
  urldate = {2024-12-10},
  chapter = {Research Article},
  copyright = {Copyright American Speech-Language-Hearing Association Apr 2019},
  langid = {english}
}

@article{DiLiberto2018CorticalMeasuresPhonemeLevel,
  title = {Cortical {{Measures}} of {{Phoneme-Level Speech Encoding Correlate}} with the {{Perceived Clarity}} of {{Natural Speech}}},
  author = {Di Liberto, Giovanni M. and Crosse, Michael J. and Lalor, Edmund C.},
  year = {2018},
  month = mar,
  journal = {eneuro},
  volume = {5},
  number = {2},
  pages = {ENEURO.0084-18.2018},
  issn = {2373-2822},
  doi = {10.1523/ENEURO.0084-18.2018},
  urldate = {2024-12-24},
  copyright = {https://creativecommons.org/licenses/by-nc-sa/4.0/},
  langid = {english}
}

@article{Mesik2021EffectsAgeCortical,
  title = {Effects of {{Age}} on {{Cortical Tracking}} of {{Word-Level Features}} of {{Continuous Competing Speech}}},
  author = {Mesik, Juraj and Ray, Lucia and Wojtczak, Magdalena},
  year = {2021},
  journal = {Frontiers in Neuroscience},
  volume = {15},
  pages = {635126},
  issn = {1662-4548},
  doi = {10.3389/fnins.2021.635126},
  langid = {english},
  pmcid = {PMC8047075},
  pmid = {33867920}
}

@article{Heinrich2016EffectiveCommunicationFundamental,
  title = {Effective Communication as a Fundamental Aspect of Active Aging and Well-Being: Paying Attention to the Challenges Older Adults Face in Noisy Environments},
  shorttitle = {Effective Communication as a Fundamental Aspect of Active Aging and Well-Being},
  author = {Heinrich, Antje and Gagne, Jean-Pierre and Viljanen, Anne and Levy, Daniel A. and {Ben-David}, Boaz and Schneider, Bruce A.},
  year = {2016},
  month = sep,
  journal = {Social Inquiry into Well-Being},
  volume = {2},
  number = {1},
  issn = {2351-6682},
  doi = {10.13165/SIIW-16-2-1-05},
  urldate = {2025-09-12},
  langid = {english}
}

@article{Helfer2017AgeRelatedChangesObjective,
  title = {Age-{{Related Changes}} in {{Objective}} and {{Subjective Speech Perception}} in {{Complex Listening Environments}}},
  author = {Helfer, Karen S. and Merchant, Gabrielle R. and Wasiuk, Peter A.},
  year = {2017},
  month = oct,
  journal = {Journal of Speech, Language, and Hearing Research},
  volume = {60},
  number = {10},
  pages = {3009--3018},
  publisher = {American Speech-Language-Hearing Association},
  doi = {10.1044/2017_JSLHR-H-17-0030},
  urldate = {2025-09-13}
}

@article{Ahissar2001SpeechComprehensionCorrelateda,
  title = {Speech Comprehension Is Correlated with Temporal Response Patterns Recorded from Auditory Cortex},
  author = {Ahissar, Ehud and Nagarajan, Srikantan and Ahissar, Merav and Protopapas, Athanassios and Mahncke, Henry and Merzenich, Michael M.},
  year = {2001},
  month = nov,
  journal = {Proceedings of the National Academy of Sciences},
  volume = {98},
  number = {23},
  pages = {13367--13372},
  publisher = {Proceedings of the National Academy of Sciences},
  doi = {10.1073/pnas.201400998},
  urldate = {2025-05-25}
}

@article{Ding2012NeuralCodingContinuous,
  title = {Neural Coding of Continuous Speech in Auditory Cortex during Monaural and Dichotic Listening},
  author = {Ding, Nai and Simon, Jonathan Z.},
  year = {2012},
  month = jan,
  journal = {Journal of Neurophysiology},
  volume = {107},
  number = {1},
  pages = {78--89},
  publisher = {American Physiological Society},
  issn = {0022-3077},
  doi = {10.1152/jn.00297.2011},
  urldate = {2025-09-14}
}

@article{Bolt2024HearingCognitiveDecline,
  title = {Hearing and Cognitive Decline in Aging Differentially Impact Neural Tracking of Context-Supported versus Random Speech across Linguistic Timescales},
  author = {Bolt, Elena and Kliestenec, Katarina and Giroud, Nathalie},
  editor = {Rothermich, Kathrin},
  year = {2024},
  month = dec,
  journal = {PLOS ONE},
  volume = {19},
  number = {12},
  pages = {e0313854},
  issn = {1932-6203},
  doi = {10.1371/journal.pone.0313854},
  urldate = {2025-01-07},
  langid = {english}
}

@article{Herrmann2022NeuralSignatureRegularity,
  title = {A Neural Signature of Regularity in Sound Is Reduced in Older Adults},
  author = {Herrmann, Bj{\"o}rn and Maess, Burkhard and Johnsrude, Ingrid S.},
  year = {2022},
  month = jan,
  journal = {Neurobiology of Aging},
  volume = {109},
  pages = {1--10},
  issn = {0197-4580},
  doi = {10.1016/j.neurobiolaging.2021.09.011},
  urldate = {2025-06-10}
}

@article{Amlerova2022EmotionalProsodyRecognition,
  title = {Emotional Prosody Recognition Is Impaired in {{Alzheimer}}'s Disease},
  author = {Amlerova, Jana and Lacz{\'o}, Jan and Nedelska, Zuzana and Lacz{\'o}, Martina and Vyhn{\'a}lek, Martin and Zhang, Bing and Sheardova, Kate{\v r}ina and Angelucci, Francesco and Andel, Ross and Hort, Jakub},
  year = {2022},
  month = apr,
  journal = {Alzheimer's Research \& Therapy},
  volume = {14},
  number = {1},
  pages = {50},
  issn = {1758-9193},
  doi = {10.1186/s13195-022-00989-7},
  urldate = {2025-09-14}
}

@article{Jiang2024ComprehensionAcousticallyDegraded,
  title = {Comprehension of Acoustically Degraded Emotional Prosody in {{Alzheimer}}'s Disease and Primary Progressive Aphasia},
  author = {Jiang, Jessica and Johnson, Jeremy C. S. and {Requena-Komuro}, Ma{\"i}-Carmen and Benhamou, Elia and Sivasathiaseelan, Harri and Chokesuwattanaskul, Anthipa and Nelson, Annabel and Nortley, Ross and Weil, Rimona S. and Volkmer, Anna and Marshall, Charles R. and Bamiou, Doris-Eva and Warren, Jason D. and Hardy, Chris J. D.},
  year = {2024},
  month = dec,
  journal = {Scientific Reports},
  volume = {14},
  number = {1},
  pages = {31332},
  publisher = {Nature Publishing Group},
  issn = {2045-2322},
  doi = {10.1038/s41598-024-82694-z},
  urldate = {2025-09-14},
  copyright = {2024 The Author(s)},
  langid = {english}
}

@article{Bidelman2017MildCognitiveImpairment,
  title = {Mild {{Cognitive Impairment Is Characterized}} by {{Deficient Brainstem}} and {{Cortical Representations}} of {{Speech}}},
  author = {Bidelman, Gavin M and Lowther, Jill E and Tak, Sunghee H and Alain, Claude},
  year = {2017},
  month = mar,
  journal = {J. Neurosci.},
  volume = {37},
  number = {13},
  pages = {3610--3620},
  issn = {0270-6474},
  doi = {10.1523/JNEUROSCI.3700-16.2017}
}

@incollection{Wong2019HongKongCantonese,
  title = {The {{Hong Kong Cantonese Corpus}}: Design and Uses},
  booktitle = {Linguistic Corpus and Corpus Linguistics in the {{Chinese}} Context},
  author = {Wong, {\relax LYM} and Luke, {\relax KK}},
  year = {2019},
  volume = {43},
  pages = {312},
  publisher = {The University of California},
  address = {Berkeley},
  isbn = {2409-2878}
}

@inproceedings{Lee2022PyCantoneseCantoneseLinguistics,
title = "{{PyCantonese: Cantonese Linguistics and NLP in Python}}",
   author = "Lee, Jackson L.  and
      Chen, Litong  and
      Lam, Charles  and
      Lau, Chaak Ming  and
      Tsui, Tsz-Him",
   booktitle = "Proceedings of The 13th Language Resources and Evaluation Conference",
   month = jun,
   year = "2022",
   publisher = "European Language Resources Association",
   language = "English",
}

@unpublished{Bates2014FittingLinearMixedEffects,
  title = {Fitting {{Linear Mixed-Effects Models}} Using Lme4},
  author = {Bates, Douglas and M{\"a}chler, Martin and Bolker, Ben and Walker, Steve},
  year = {2014},
  month = jun,
  journal = {arXiv [stat.CO]},
  isbn = {1406.5823}
}

@manual{Lenth2025EmmeansEstimatedMarginal,
  type = {Manual},
  title = {Emmeans: {{Estimated}} Marginal Means, Aka Least-Squares Means},
  author = {Lenth, Russell V.},
  year = {2025}
}

@inproceedings{So2025PerformanceMontrealForced,
  title = {Performance of {{Montreal Forced Aligner}} on {{Cantonese Spontaneous Speech}}},
  booktitle = {Proc. {{Interspeech}} 2025},
  author = {So, Ka Ki and Xu, Chenzi and Cao, Grace Wenling and Mok, Peggy},
  year = {2025},
  pages = {5398--5402},
  doi = {10.21437/Interspeech.2025-1882},
  urldate = {2025-08-26},
  langid = {english}
}

@article{Fong2021CanInhibitionDeficit,
  title = {Can Inhibition Deficit Hypothesis Account for Age-Related Differences in Semantic Fluency? {{Converging}} Evidence from {{Stroop}} Color and Word Test and an {{ERP}} Flanker Task},
  author = {Fong, Manson Cheuk-Man and Law, Tammy Sheung-Ting and Ma, Matthew King-Hang and Hui, Nga Yan and Wang, William Shiyuan},
  year = {2021},
  month = jul,
  journal = {Brain and Language},
  volume = {218},
  pages = {104952},
  issn = {0093-934X},
  doi = {10.1016/j.bandl.2021.104952}
}

@article{Ma2021RegularityRandomnessAgeing,
  title = {Regularity and Randomness in Ageing: {{Differences}} in Resting-State {{EEG}} Complexity Measured by Largest {{Lyapunov}} Exponent},
  author = {Ma, Matthew King-Hang and Fong, Manson Cheuk-Man and Xie, Chenwei and Lee, Tan and Chen, Guanrong and Wang, William Shiyuan},
  year = {2021},
  month = dec,
  journal = {Neuroimage: Reports},
  volume = {1},
  number = {4},
  pages = {100054},
  issn = {2666-9560},
  doi = {10.1016/j.ynirp.2021.100054}
}

@article{Wang1973ChineseLanguage,
  title = {The {{Chinese Language}}},
  author = {Wang, William S-Y.},
  year = {1973},
  month = feb,
  journal = {Scientific American},
  volume = {228},
  number = {2},
  pages = {50--60},
  issn = {0036-8733},
  doi = {10.1038/scientificamerican0273-50},
  urldate = {2025-02-18}
}

@article{Mixdorff2017AudiovisualExpressionsAttitude,
  title = {Audio-Visual Expressions of Attitude: {{How}} Many Different Attitudes Can Perceivers Decode?},
  shorttitle = {Audio-Visual Expressions of Attitude},
  author = {Mixdorff, Hansj{\"o}rg and H{\"o}nemann, Angelika and Rilliard, Albert and Lee, Tan and Ma, Matthew K.H.},
  year = {2017},
  month = dec,
  journal = {Speech Communication},
  volume = {95},
  pages = {114--126},
  issn = {01676393},
  doi = {10.1016/j.specom.2017.08.009},
  urldate = {2025-09-18},
  langid = {english}
}

@article{Degano2024SpeechProsodyEnhances,
  title = {Speech Prosody Enhances the Neural Processing of Syntax},
  author = {Degano, Giulio and Donhauser, Peter W. and Gwilliams, Laura and Merlo, Paola and Golestani, Narly},
  year = 2024,
  month = jun,
  journal = {Communications Biology},
  volume = {7},
  number = {1},
  pages = {1--10},
  publisher = {Nature Publishing Group},
  issn = {2399-3642},
  doi = {10.1038/s42003-024-06444-7},
  urldate = {2024-08-23},
  copyright = {2024 The Author(s)},
  langid = {english}
}

@article{Kovacs2023SpeechProsodySupports,
  title = {Speech Prosody Supports Speaker Selection and Auditory Stream Segregation in a Multi-Talker Situation},
  author = {Kov{\'a}cs, Petra and T{\'o}th, Brigitta and Honbolyg{\'o}, Ferenc and Szal{\'a}rdy, Orsolya and Koh{\'a}ri, Anna and M{\'a}dy, Katalin and Magyari, Lilla and Winkler, Istv{\'a}n},
  year = 2023,
  month = apr,
  journal = {Brain Research},
  volume = {1805},
  pages = {148246},
  issn = {0006-8993},
  doi = {10.1016/j.brainres.2023.148246},
  urldate = {2026-02-27}
}

@article{Gillis2023NeuralTrackingLinguistic,
  title = {Neural Tracking of Linguistic and Acoustic Speech Representations Decreases with Advancing Age},
  author = {Gillis, Marlies and Kries, Jill and Vandermosten, Maaike and Francart, Tom},
  year = 2023,
  month = feb,
  journal = {NeuroImage},
  volume = {267},
  pages = {119841},
  issn = {1053-8119},
  doi = {10.1016/j.neuroimage.2022.119841},
  urldate = {2026-02-27}
}

@article{Gnanateja2025CorticalProcessingDiscrete,
  title = {Cortical Processing of Discrete Prosodic Patterns in Continuous Speech},
  author = {Gnanateja, G. Nike and Rupp, Kyle and Llanos, Fernando and Hect, Jasmine and German, James S. and Teichert, Tobias and Abel, Taylor J. and Chandrasekaran, Bharath},
  year = 2025,
  month = mar,
  journal = {Nature Communications},
  volume = {16},
  number = {1},
  pages = {1947},
  publisher = {Nature Publishing Group},
  issn = {2041-1723},
  doi = {10.1038/s41467-025-56779-w},
  urldate = {2026-01-17},
  copyright = {2025 The Author(s)},
  langid = {english}
}

@article{Goldstein2022SharedComputationalPrinciples,
  title = {Shared Computational Principles for Language Processing in Humans and Deep Language Models},
  author = {Goldstein, Ariel and Zada, Zaid and Buchnik, Eliav and Schain, Mariano and Price, Amy and Aubrey, Bobbi and Nastase, Samuel A and Feder, Amir and Emanuel, Dotan and Cohen, Alon and Jansen, Aren and Gazula, Harshvardhan and Choe, Gina and Rao, Aditi and Kim, Catherine and Casto, Colton and Fanda, Lora and Doyle, Werner and Friedman, Daniel and Dugan, Patricia and Melloni, Lucia and Reichart, Roi and Devore, Sasha and Flinker, Adeen and Hasenfratz, Liat and Levy, Omer and Hassidim, Avinatan and Brenner, Michael and Matias, Yossi and Norman, Kenneth A and Devinsky, Orrin and Hasson, Uri},
  year = 2022,
  month = mar,
  journal = {Nature Neuroscience},
  volume = {25},
  number = {3},
  pages = {369--380},
  publisher = {Nature Publishing Group},
  issn = {1097-6256},
  doi = {10.1038/s41593-022-01026-4},
  urldate = {2022-09-15}
}

@software{neurothew_2026_20748010,
  author       = {neurothew},
  title        = {neurothew/cantonese-phonotactic-features: v1.0.0},
  month        = jun,
  year         = 2026,
  publisher    = {Zenodo},
  version      = {v1.0.0},
  doi          = {10.5281/zenodo.20748010},
  url          = {https://doi.org/10.5281/zenodo.20748010},
  swhid        = {swh:1:dir:ae9b7749d2d0eaac7459ca4c55729e7d805bfc4d
                   ;origin=https://doi.org/10.5281/zenodo.20748009;vi
                   sit=swh:1:snp:07d6fdf1cc788e5f434dab5a4512c1dc338a
                   57e5;anchor=swh:1:rel:a176904bfb536e4e192d3083a31c
                   e862fc6a13fe;path=neurothew-cantonese-phonotactic-
                   features-e4999af
                  },
}

\end{document}